\RequirePackage{fix-cm}
\documentclass[smallextended]{svjour3-arxiv}       
\smartqed  
\usepackage{graphicx}
 \usepackage{sidecap} 
\usepackage{graphicx,amsmath,amssymb,amsbsy,latexsym,verbatim}
\usepackage{makeidx}
\usepackage{color}

\newcommand{\be}{\nopagebreak[3]\begin{equation}}
\newcommand{\ee}{\end{equation}}
\newcommand{\ba}{\nopagebreak[3]\begin{eqnarray}}
\newcommand{\ea}{\end{eqnarray}}

\newcommand{\bc}{}
\newcommand{\C}{\mathbb{C}}

\journalname{.. }
\begin{document}

\title{Measuring the last burst of non-singular black holes
}


\author{Francesca Vidotto         
}


\institute{
University of the Basque Country UPV/EHU, Departamento de F\'isica Te\'orica,\\
Facultad de Ciencia y Tecnolog\'ia, 
Barrio Sarriena s/n, 48940 Leioa, Bizkaia, Spain\\
              \email{francesca.vidotto@ehu.es}           
}

\date{May 9, 2017}

\maketitle

\begin{abstract}
Non-perturbative quantum gravity prevents the formation of curvature singularities and may allow black holes to decay with a lifetime shorter than evaporation time. This, in connection with the existence of primordial black holes, could open a new window for quantum-gravity phenomenology. I discuss the possibility of observing astrophysical emissions from the explosion of old black holes in the radio and in the gamma wavelengths. These emissions can be discriminated from other astrophysical sources because of a peculiar way the emitted wavelength scales with the distance. The spectrum of the diffuse radiation produced by those objects presents a peculiar distortion due to this scaling.
\end{abstract}

\section{Historical Preamble}

George Lema\"itre left many legacies, well beyond his technical results in cosmology. Three of them still  riverberate on the current research in quantum gravity.
\begin{description}
\item[\it Finiteness] ~~ A finite universe is a recurring theme in Lema\"itre research. This is clear in his best known idea: a universe finite in time, i.e. having a beginning rather than being eternal as the static universe, which was the dominant paradigm at the time. But finiteness inspired many of his other ideas. His ideas about the origin of the universe in the form of a quantum \emph{primeval atom} addressed the plague of curvature singularities in General Relativity. The quantum properties of the primeval atom were supposed to correct for this, even if at the time no available technique could achieve it. Less known is the fact that George Lema\"itre addressed the problem of singularities also in the context of black holes during his doctorate: he analysed a non-singular black-hole solution with non-constant density, that lead later to 
the Eddington-Lema\"itre metric for cosmology.
\item[\it Phoenix Universe] ~~ 
The Phoenix Universe is a model where the universe goes trough oscillations, collapsing and bouncing back. It reestablishes infiniteness in time, but avoids the problem of the initial curvature singularity. The idea of a universe that raises from its own ashes was welcomed by many as conceptually preferable,  and dismissed by others (Eddington notoriously said \emph{``I am no Phoenix worshipper''}). Lema\"itre was attracted by the by the philosophical aspects of his idea, but renounced it when his calculations  yield cosmological cycles that were too short. Today viable bouncing models are studied, relying of different physical mechanisms \cite{Brandenberger2016}. The cosmological framework based on Loop Quantum Gravity realises this possibility via a quantum singularity resolution, as I briefly illustrate in the next Section.
\item[\it Phenomenology] ~~ 
George Lema\"itre understood that data were indicating that the universe was expanding. He was the first to write the law that became later known as the Hubble law: not only this was in 1927, two years before 1929 Hubble's paper, but, most importantly, he had already the correct interpretation of the data. His attention for data was stronger than for most of his contemporary pears working in gravitation ---Einstein included. Theory needs data!  Today, in our era of celebrated data abundance in cosmology, our attention should perhaps shift towards the theory to find the right interpretation, without which the accumulation of new data remains blind. I also like to mention here the pioneeringh work of Lem\"itre in  numerical simulations. Simulations are the theorist's experiment, and in quantum gravity simulations are going to play an increasingly important role guiding the future analytical investigation 
\cite{Magliaro:2008dq,Christensen:2009oq,Bahr2015,Bayle2016,Dona2018,Gozzini2018}%
.
\end{description}

\section{Quantum Singularity Resolution}
The singularities of General Relativity can be avoided if matter violates the strong energy condition which is an hypothesis in the Penrose-Hawking singularity theorem. They can also be avoided quantum-mechanically: this is what is expected in a viable quantum gravity theory, and this is what happens in Loop Quantum Gravity.

Loop Quantum Gravity is based on gauge-theory perspective on General Relativity, written in the tetrad formalism. On a fixed time surface, tetrads reduce to triads, and the corresponding operators in the quantum theory are generators of rotations. These are the basic ingredients to define geometric operators. The compactness of the rotation group (or its double covering $SU(2)$) yields the discreteness of the spectrum of area and volume operators, which is the cornerstone of Loop Quantum Gravity. This result alone, however, is not sufficient to account for the  resolution of the singularity of classical General Relativity. This depends on dynamics of the theory.

A covariant formulation of the Loop Quantum Gravity dynamics is given by the \emph{spinfoam} formalism. In this formalism the geometrical $SU(2)$ quantities labelling the states at a given time evolve building up  Lorentzian spacetime. The corresponding transition amplitudes are defined by a map from 
representations of  $SU(2)$ to unitary representations of $SL(2,\C)$. This map induces a proportionality $\vec K=\gamma \vec L$ between the generators of rotations $\vec L$ and the generator of boosts $\vec K$ in $SL(2,\C)$%
\footnote{This relations is called the \emph{simplicity} constraint. Starting from the topological BF action, this constraint frees the gravitational degrees of freedom and project into the physical states, encoding the dynamics of spacetime \cite{Rovelli:2014ssa}.}.
The proportionality constant $\gamma$ is called after Barbero and Immirzi: in the classical tetrad formulations it multiplies the Holst term in the action%
\footnote{The Holst term appearing in General Relativity is analogue to the other one we know for a non-abelian theory, the $\theta$-term in QED: these terms do not affect the classical equations of motion, but they play a role in the quantum theory.}.
The equation $\vec K=\gamma \vec L$ ties the discrete spectrum of $L^2$ to the discretness of the  boost spectrum and this, in turn, implies the existence of a maximal acceleration \cite{Rovelli2013a}. A maximal acceleration have been considered in the study of singularities on a classical spacetime: having a maximal acceleration appears to be a sufficient condition to have a bounded curvature and therefore no curvature singularity%
\footnote{The definition of a curvature singularity in General Relativity is more cumbersome than just checking the finiteness of the curvature. Indeed, one should check also that spacetime is geodesically complete. Given a maximal acceleration, there is one (only) counterexample of non-geodesically complete spacetime given by Tipler \cite{Tipler:1977kx}, but it require a quite artificial construction that does not apply to the kind of spacetime we may want to consider in physical situations.}%
. Therefore, in the covariant formulation of Loop Quantum Gravity, there are the conditions for a quantum \emph{no-singularity comnjecture}.

Curvature singularities are therefore expected to disappear in any formulation of a quantum cosmology based on the loop quantization.  An extensive literature has confirmed this expectation in the context of Loop Quantum Cosmology \cite{Agullo2017}. In Loop Quantum Cosmology a quantum Hamiltonian operator is constructed using holonomies of the Ashtekar connection, which codes the curvature. Promoting holonomies directly to quantum operators yields boundness. In this simplified framework one can study the consequences of this boundness explicitly: the effective equations of motion for the  universe, show that when the universe approaches a Planckian energy density, quantum effects manifest as an effective repulsive force that prevent further collapse. In a (closed) primordial universe, this energy density is attained when the universe is many orders of magnitude larger than a Planck radius \cite{Ashtekar:2006es}. The result is a bouncing universe: the universe experiences a pre-bigbang contracting phase, the contraction is stopped by the appearance of a quantum repulsive force, that lead to the subsequent expanding phase.

In the rest of the paper I focus on the another main consequence of the quantum singularity resolution: its effect on black holes. 

\section{Planck Stars}

It is reasonable to expect the physics of the singularity at the center of a black hole to be similar to that of the cosmological bounce.  The application of Loop Quantum Gravity to the central black-hole singularity started a decade ago \cite{Modesto2004,Modesto2006,Bojowald:2005qw,Ashtekar2005,Gambini:2008bh} and is actively continuing today \cite{Gambini2014,Gambini2013,Gambini2015,Gambini2016,Corichi2016,Yonika2017,Olmedo2017}.  A  detailed picture recently developed is provided by the \emph{Planck Star} scenario: a distribution of matter collapses, forms a black hole and continues the collapse inside its dynamical horizon, the quantum repulsion prevents the formation of the central singularity and triggers a phase of expansion, eventually muting the trapping surface of the horizon into an anti-trapping surface, i.e. a white hole \cite{Rovelli2014h}. The name Planck Star comes from the analogy with usual stars because the quantum repulsion prevents a further collapse like the nuclear reactions do in the core of stars. Notice that the collapse still produce an horizon, but a dynamical horizon rather than an event horizon, therefore most of the usual black holes physics is unchanged.   

The Planck star scenario bears three novelties. First, the black hole bounces back in the same universe and in the same spacial point where it is, and not to a different region like for instance in the case of an Einstein-Rosen bridge. Previous attempt to bridge a collapsing solution with an expanding one \cite{Stephens1994,Frolov1981} faced the problem that in Penrose's conformal diagrams a white hole sit in the causal past of a black hole, not in its future. The realisation that it is possible to have a white hole in the future of a black hole, with the same asymptotic region, was clarified by Haggard and Rovelli \cite{Haggard:2014fv}. Such a spacetime is forbidden by the classical Einstein equations, but is possible if these are violated in a small spacetime region, as typical of quantum tunnelling.

The second feature of the Planck Star framework is that it is possible to ask how long the bounce of a black hole takes.  Notice that the bounce can be very fast in the proper time comoving with the collapsing matter ---of the order of the time light would take to cross the horizon from one side to the other--- but very long for an external observer, i.e. an observer that see a front of matter collapsing into forming the black hole and then coming back from a white hole. The difference can be huge, due to the large redshift due to the gravitational potential. The external bounce time  is the time we are interested to study in view of the phenomenology; this is the black hole lifetime.

In the covariant theory of Loop Quantum Gravity, it is possible in principle to  compute the black hole lifetime \cite{Christodoulou:2016ve} althought the explicit calculation has so far proven cumbersome \cite{Christodoulou2018}.  The same quantity should be understandable using different quantum-gravity approaches and a concrete result on that may provide a test for the theory. From higher-dimensional black holes to a brane interior, there is a convergence toward a possible instability of quantum-gravitational origin that can yields an explosive event \cite{Gregory:1993vy,Casadio:2000py,Casadio:2001dc,Emparan:2002jp,Kol:2004pn}. 

In the lack of a solid explicit computation, it is possible to estimate the allowed lifetime using heuristic arguments. In the simplified case of a non-rotating black hole with no charge, this time would depend solely on the total mass. Therefore, we can express it in natural units as a function of the black hole mass $M_{BH}$. 

The characteristic time scale of the Hawking evaporation, is $M_{BH}^3$. The usual computation of Hawking evaporation uses a fixed background and its validity should be questioned when the dynamical nature of spacetime needs to be taken into account. This is expected to happen when approximately half of the mass of the black hole has evaporated, as signalled by the inconsistency referred as `firewall' \cite{Almheiri:2012rt} found in this regime. Therefore a full quantum-gravity regime should begin before firewalls would appear. The time $M_{BH}^3$ can be estimated as the longest possible time the hole can live before turning into a white-hole with the consequent explosion.

To bound the black hole lifetime time from below, consider that quantum fluctuations can already appear on a shorter time scale, when Hawking radiation is still negligible. In general a classical system present a \emph{quantum-break time}  that corresponds to its characteristic timescale after which the evolution departs from classicality and the system should be treated as a quantum system. This time can be computed for a black hole asking how long would it take for quantum fluctuations to be detected  outside the horizon. The curvature at the horizon could be small (the bigger the black hole the smaller this curvature would be) but never zero. As this curvature is proportional to $M_{\rm BH}^{-2}$ in the area outside the horizon, the  inverse of this quantity gives a time $M_{\rm BH}^2$ \cite{Haggard:2014fv,Haggard:2016ibp}, a time hugely shorter than $M_{\rm BH}^3$ as one would realise by plugging the Planck constants in the equation.   $M_{\rm BH}^2$ is the lower bound on the possible lifetime of a hole: in order for a black hole to decay into a white hole, some quantum effects should appear and $M_{\rm BH}^2$ is the minimal time for which this could happen. In the treatment of quantum system that undergo a decay, the distribution of events is strongly peaked on the minimal time. This suggest that for black holes the time $M_{\rm BH}^2$ might be the relevant one.

Finally, the third novel aspect of the Planck-Star scenario is a whole new phenomenology of the explosive expanding phase. The current studies, while motivated by Planck Stars, can in fact apply to a wider range of model of exploding black holes. I discuss this in the remaining part of this presentation, considering the full range of possible black-hole lifetimes between $M_{\rm BH}^2$ and $M_{\rm BH}^3$. 

It is convenient to parametrise this window as a lifetime
$$
\tau = 4\kappa\left({M_{\rm BH}}/{m_{\rm Pl}}\right)^{2} t_{\rm Pl} 
$$ where $\kappa$ is a phenomenological parameter that for primordial black hole exploding now
ranges over the wide inteval $0.05$ to $10^{22}$. We use Planck units for time $t_{\rm Pl } $ and mass $ m_{\rm Pl } $.
The lifetime determines the size of the hole at the time of its explosion. 
Let's estimate the size of a black hole that has formed in the early universe and has lived as long as the Hubble time $t_H \approx 14 \cdot 10^9$years. Then within the running of $\kappa$, the mass of this black hole would range between $10^{14}g$ and $10^{23}g$, that yields approximatively a Schwarzschild radius respectively between $10^{-14}cm$ and $10^{-2}cm$ \cite{Rovelli2014h,Barrau2014g}.
Given the mass of the black hole at the time of the explosion, we can study the characteristics of the astrophysical signal, both via gravitational and electromagnetic messengers. I will not consider here the emission in gravitational waves. I focus instead on the electromagnetic one.

\section{Expected Astrophysical Signals}
We do not know the exact mechanism of emission from a white hole. Still, the possible signals arising from such an explosion can be described and searched. Three possible astrophysical signals can be expected. These are: a low energy signal that depends on the size of the hole exploding and whose wavelength varies widely within the allowed lifetime window;  a high-energy signal in the very high energy spectral band that depends on the energy of the matter that formed the black hole;  and a signal in the radio that would be produced if there are magnetic fields in the surrounding of the exploding hole. I discuss the three cases below. 

\subsection{Low Energy Channel} \label{LEC}
During an explosion, the exploding object have its modes excited. The lowest modes are  associated with the size of the object. For an exploding black hole, we can expect a mode of the order of the Schwarzschild radius to be excited and appear  as a channel in the emitted signal. This should not be surprising as a similar reasoning yields to the wavelength of the particle produced by Hawking radiation, where the wavelength is peaked around the Schwarzschild radius with a factor $\sim16$ that accounts for a combination of physical constants.

Black holes formed by stellar collapse are not of interest in this context, because they would explode only in the far future. Only \emph{primordial black holes} (PBH) formed in the early universe may have the size and the lifetime to have already exploded and be within reach for observation. Different mechanisms can lead to the formation of black holes in the very early universe: in fact, there are so many mechanisms, ranging from versions of inflation to cosmic strings, that their existence seems quite plausible. 
Let's consider a black hole whose lifetime is equal to the Hubble time and use the relation for the lifetime $\tau = 4\kappa\left({M_{\rm BH}}/{m_{\rm Pl}}\right)^{2} t_{\rm Pl}$. 

We have studied in \cite{Barrau:2015uca} the emission for all possible value of $\kappa$ and for the highest one, corresponding to longer lifetimes, the emitted signal is in the $GeV$. Studying with the PYTHIA code \cite{Sjostrand:2014zea} the spectrum of photons produced in the event, it has been shown that the highest photon density would be in the $MeV$, making a detection more likely in this band \cite{Barrau2014e}. Short Gamma Ray Bursts \cite{Nakar:2007yr}, not associated with known sources or with possible neutron-star merging, would candidate as such signal. 

For the lowest values of $\kappa$, favoured by the theory, we can expect a signal reaching 1imatively the half millimetre. In this wavelength, events within all the visible universe would be detectable \cite{Barrau:2015uca}. Instruments like the Large Latin American Millimeter Array (LLAMA) could be promising for the detection.

An aspect that makes this channel with this value of $\kappa$ particularly interesting is a possible connection with Fast Radio Bursts (FRB) \cite{Barrau2014g}. The bursts present a number of characteristics that would fit our signal, such as the (mostly) extragalactic origin, the unusually high flux, the rate of events, and so on. Indeed, there is a mismatch between our prediction and the wavelength of FRB, that is around $20\,cm$. If one wants to further investigate this possibility, there are two possible explanation for this mismatch. One is simply given by the rough simplifications taken in the calculation, such that they can easily lead to miss a couple of orders of magnitude. A more interesting possibility considers the random nature of black-hole lifetime. Therefore, FRB could be just part of a family of cosmic rays coming from black holes explosions, of which those in the radio are selected due to the atmosphere opacity and the telescopes availability. 
This possibility becomes more viable if the PBH mass spectrum is highly peaked \cite{Barrau2018}.

\subsection{High Energy Channel}\label{HEC}
From the point of view of the matter collapsing and forming the black hole, the bouncing process is extremely rapid and almost elastic: it is conceivable that no dissipative effect affects appreciably the matter energy. Therefore, matter could be expelled with approximately the same energy with which the process started.

We are interest in black holes formed in the early universe.
The first mechanism proposed for their formation \cite{Carr:1974nx}, and still the most palatable, is based on the presence of overdense regions during the reheating: their collapse forms PBHs whose size depends on the Hubble horizon at the precise time of their formation. The universe then would be composed mostly by photons with a temperature of the order of $TeV$. Correspondently, this is the energy we expect for the matter emitted by a white hole. Being in the $TeV$, we label this emission channel as the ``high energy channel''. 

A direct observation of a cosmic ray of such an high energy has an observational horizon due to the interaction with the CMB. As studied in \cite{Barrau:2015uca}, only events within our galactic neighbourhood are at reach. This estimation is performed taking into account the observational facilities currently available. On the other hand, new telescopes will start to operate in the forthcoming years \cite{Rieger:2013awa} allowing for the exploration of this observational window. 

\subsection{Rees Mechanism}\label{RM}
The particle production in the high energy channel can be studied using codes like PYTHIA \cite{Sjostrand:2014}. It would be important to check in more detail the coherence of the signal and the production of electron-position pairs. The number of these pairs is expected to be significant, as in the explosive process all the mass of the black hole is emitted. The resulting relativistic expanding shell behaves as a perfect conductor. In the likely presence of interstellar magnetic fields, the shell would reflect and boost the virtual photons of those fields. The resulting emission depends on the Lorentz factor of the expanding shell, that depends on the energy of the initial event. As in this case the energy is in the $TeV$, this allow to efficiently convert the primary signal into a secondary one in the radio. The spectrum of this emission corresponds to the one derived by Blandford \cite{Blandford1977}. Rees in the Seventies studied this mechanism \cite{Rees1977} in the hope of observing the  explosion of PBH: a signal in the radio can travel without deteriorating interactions much further than one at the $TeV$ energy, at which the explosion were expected. Today we join Rees in that hope when considering the possibility that the detection of Fast Radio Bursts could be in fact the detection of PBH explosions.

The connection between the Rees mechanics and PBH exploding in a time shorter than the evaporation time have been considered in \cite{Kavic2008,Kavic2008a,Estes2017}. 
The connection between the Rees mechanics and FRB, on the other hand, is consider as well by Thompson \cite{Thompson2017}, even though with a different exploding source.
The Planck Star framework allows to pull together these lines of research, with a better defined theoretical framework for the explosion and, as mentioned before, the right characteristic in terms of localisation mostly out of the galaxy, the efficient conversion of the black hole mass in the received flux, and the expected rate of events.

\section{A quantum signature}
The most interesting aspect of the Planck Star scenario is the fact that the signal has a clear signature that can distinguish it from other astrophysical signals.  This is a peculiar relation between distance and observed wavelength, and it depends on the relation between the lifetime of the black hole and its mass. Consider two PBH of different mass, one of which explodes now near us and the other far away. The second one must be smaller because it exploded earlier, hence has a shorter lifetime. Notice that the smaller black hole produces a signal of the shorter wavelength \emph{both} in the low and in the high energy channels. The shorter wavelength partially compensates the standard cosmological redshift. 

\begin{SCfigure}
\includegraphics[height=55mm]{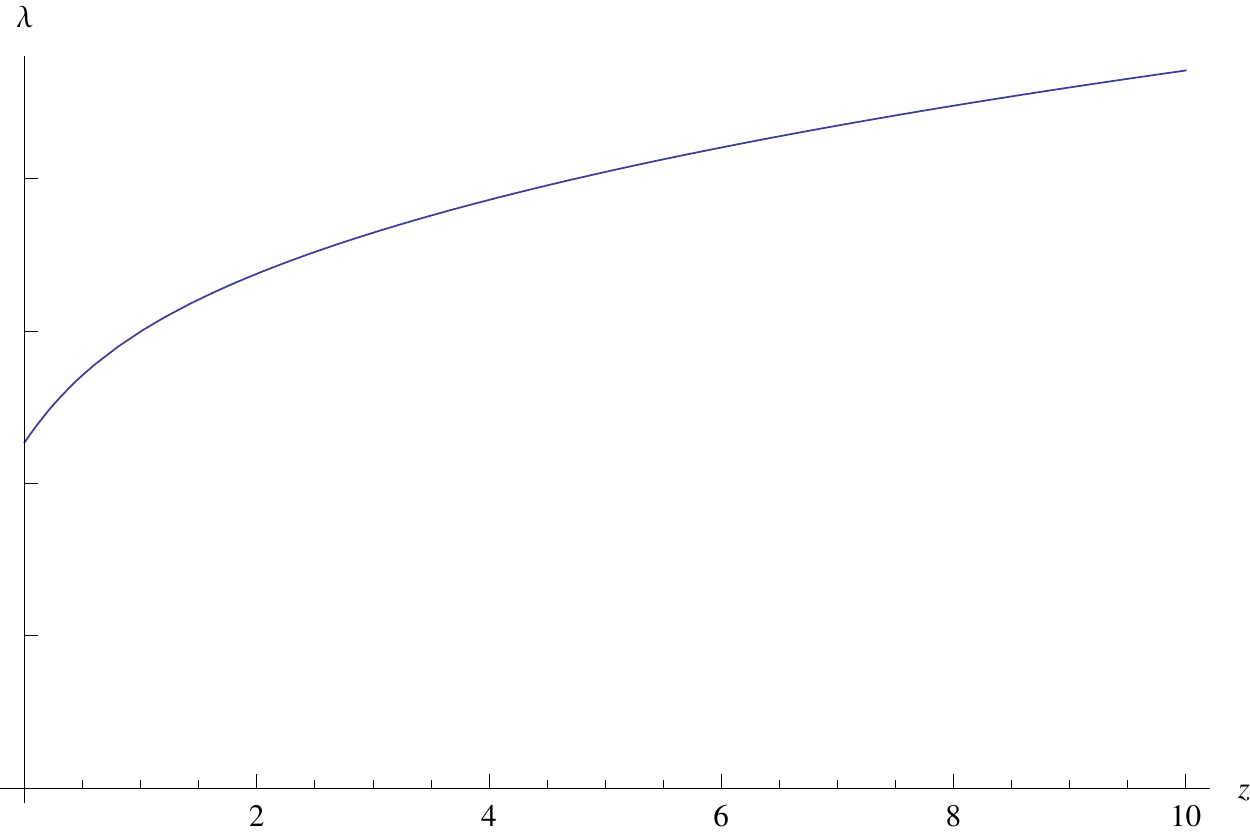}
\caption{
The observed wavelength (unspecified units) from black hole explosions as a function of the redshift $z$. The curve presents a characteristic flattening as the redshift is compensated by the shortening of the emitted wavelength for more distant sources.}
\label{fig:flat}
\end{SCfigure}

This compensation is peculiar: the emitted wavelength of a standard astrophysical object scales linearly with the distance because of the redshift. The phenomenon is not a generic feature of PBH either, as an explosion due to the Hawking evaporation happens in the standard theory when the black hole reaches the Planck size, independently from its initial mass, and the signal emitted would only know of that scale. In the case of Planck stars, on the other hand, there is a modified relation between the wavelength of the observed signal and the distance of its source \cite{Barrau:2014yka}. The resulting curve in Fig.\ref{fig:flat} presents a characterising flattening. This provide a signature of the quantum-gravitational origin of the bursts.
To fit such a curve, we need to collect data of bursts and be able to associate the source distance, a task not always possible. A standard method would utilise the dispersion measure of the received signal. In lucky situations, the burst may be associated with a known astronomical object, for instance a host galaxy. 

This framework provide a further technique to measure distances. In fact we can reverse the previous argument and use the fact that distant explosions of PBH come from less massive ones. Therefore, the incoming flux will be lower. This is particularly interesting in connection with a FRB: in fact, the Rees mechanism washes the information about the size of the source in terms of wavelenght, even though not completely, and make more cumbersome to localise the source. Still, the flux in the radio emission will depend on the flax of the primary $TeV$ burst. Therefore, we expect to see FRB with a smaller flux the more distant is the source.

\subsection{Intensity mapping}
\begin{figure}[h]
        \includegraphics[width=.5\textwidth]{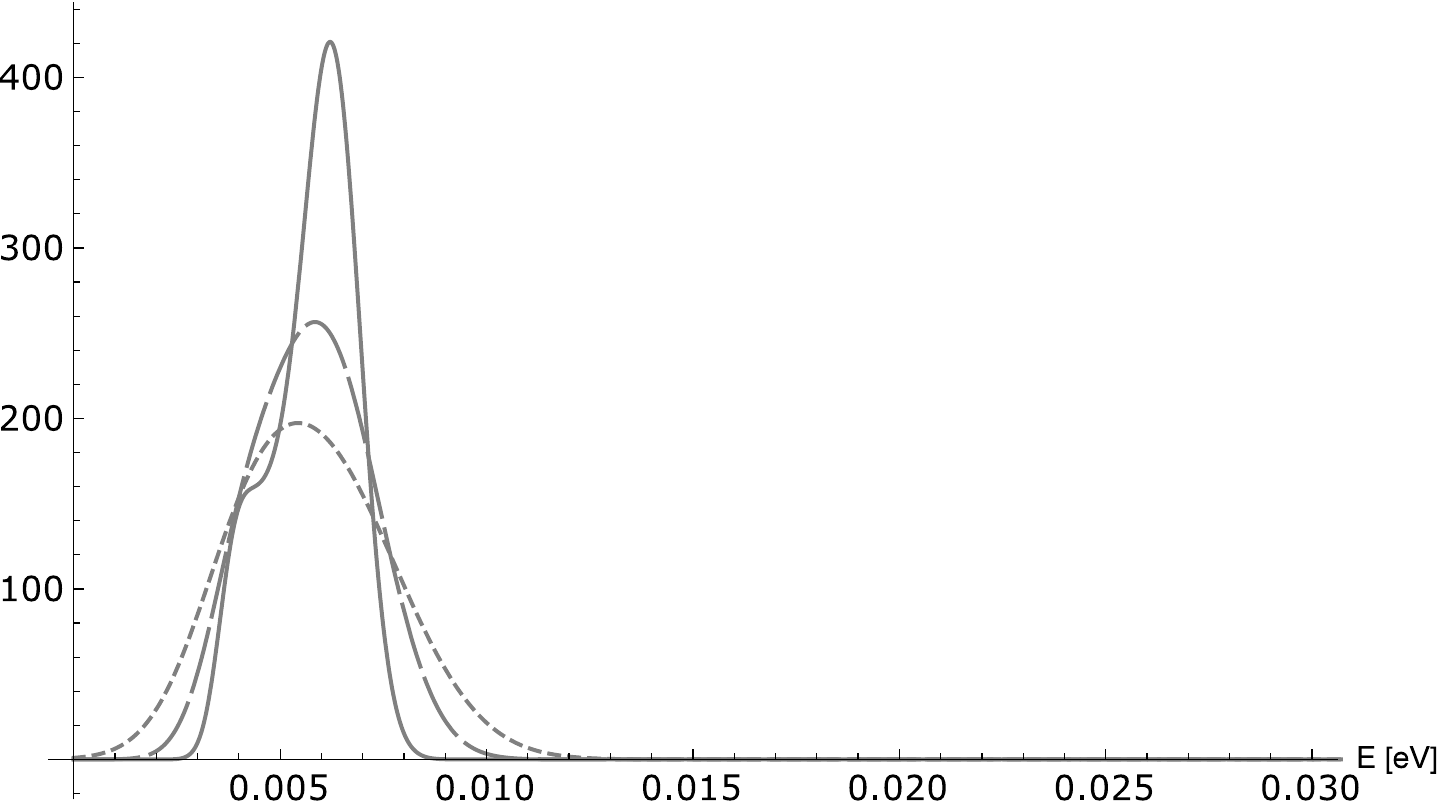}
        \includegraphics[width=.5\textwidth]{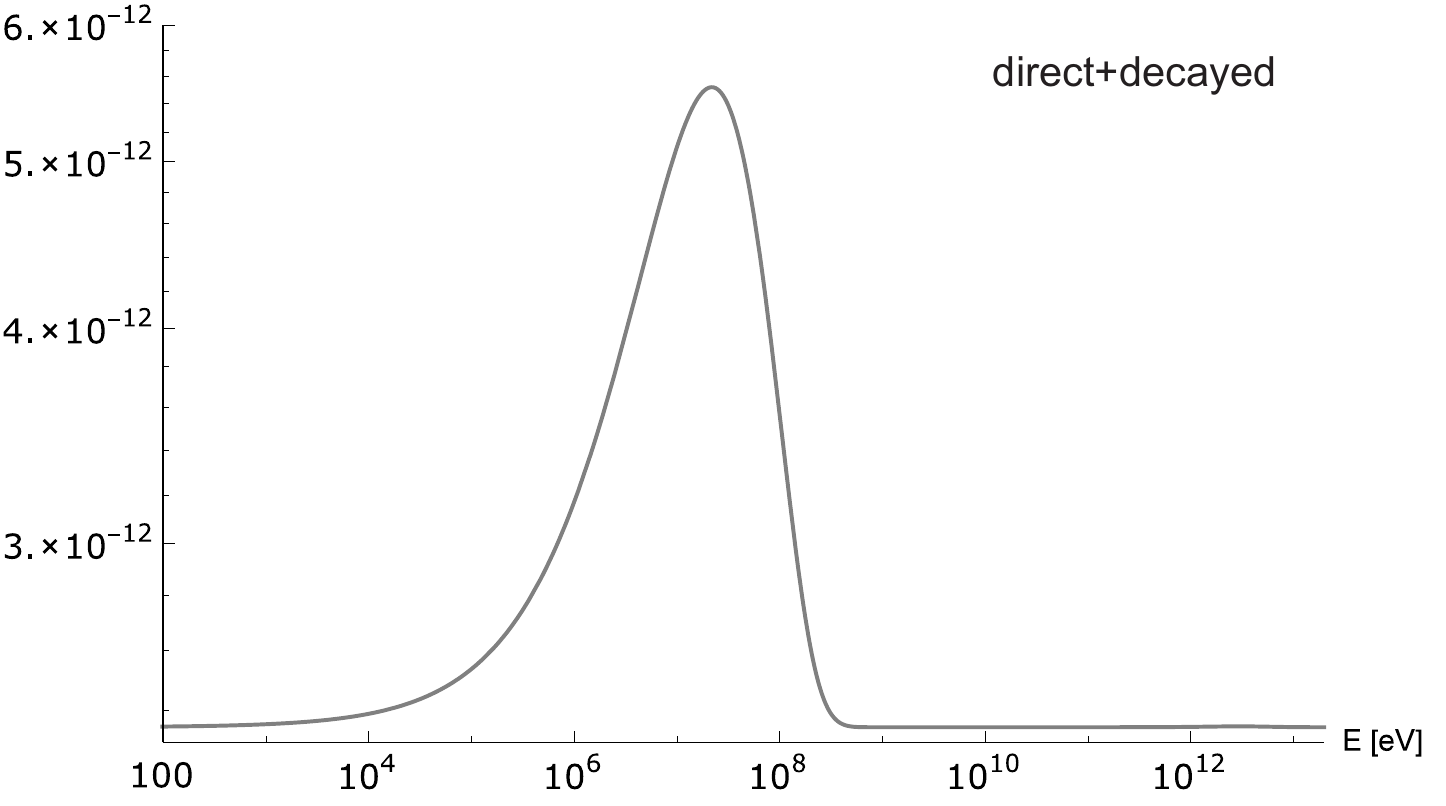}
\caption{The diffuse emission of the low energy channel (left) and the diffuse emission of the high energy channel (right) for the shortest BH lifetime ($\kappa=.05$). We have not specified the units in the ordinate axis as the normalisation of the spectrum can possibly depend on the percentage of PBH as dark matter.}
\label{fig:diffuse}
\end{figure}

Detection of cosmic rays with a precise association of the distance of their source is hard. A more affordable strategy is to set a campaign of observation on the largest sky field in order to collect the emission integrated over all the detectable past PBH explosions.  This has been studied exploring the whole range of lifetime varying the parameter $\kappa$. The single events have been simulated with the PYTHIA code and then integrated. Luckily, the characteristic wavelength-distance relation discussed before propagates its effects also in the spectrum of the integrated emission \cite{Barrau:2015uca}. In fact, the integration over standard astrophysical sources, with a given emission spectrum, produce a thermal black-body radiation. In this case instead the integrated spectrum obtained results distorted because of the peculiar redshift-wavelength relation of Fig.\ref{fig:flat}. The same effect manifests in the spectrum obtained for the high energy and the low energy channels. This result, verified for the whole window of allowed lifetimes, is here shown for the case of the shortest lifetime in Fig.\ref{fig:diffuse}. 
Notice that while the result depends on the PBH mass spectrum, a study considering different shapes for the PBH mass spectrum has shown that this has only a small effect on the shape of the emission spectrum and does not change the qualitative result  \cite{Barrau:2015uca}.

\subsection{Primordial Black Holes}
We have considered black holes forming in the very early universe, so that the matter forming them is not subject to the constraints on baryons that apply for nucleosynthesis. This allow to qualify PBHs as non-baryonic dark matter (DM). PBHs were indeed one of the first candidates historically considered to explain dark matter \cite{Chapline:1975cr}. They have had alternate fortune. Today, with the lack of direct detection of DM particles and the recent discovery of black holes of unexpected sizes, they are receiving renewed attention. We have now observational constraints for different mass ranges  of PBHs \cite{Carr:2016hva,Chen:2016pud,Green:2016xgy}. Some of the old constraints are today relaxed by allowing different mechanisms of PBH formations: most of the constraints were in fact obtained with a monochromatic mass spectrum, a paradigm now challenged by different authors as unrealistic \cite{Carr:2017jsz}. 
The framework presented here requires to reconsider farther certain of these constrains, especially those based on solely Hawking evaporation, and presents some novel phenomenological feature.

The process that yields the final explosion of a black hole can be seen as a decaying process. Therefore PBHs would qualify as a decaying DM with a distinctive characteristic: the decay time depends on the mass. This feature distinguishes this model from other decaying DM candidates, whose decay time is fixed. In particular, the decaying time $\tau$ is shorter than the evaporation time: this may reduce or even completely suppress the presence of Hawking radiation, undetermine PBH constraints based on it.
   
Another constrain challenged by this scenario concerns small black hole and their lack of detection by microlensing \cite{griest2013new}. In fact, according to Hawking evaporation, PBHs having completed their full evaporation today have a mass of $10^{12}$ Kg. All the black holes with a mass smaller than that would not be present in the universe any more, having already evaporated. In the new scenario the minimal mass of PBH still present today raises. For the shorter lifetime $M_{\rm BH}^2$, PBHs decaying and exploding today have a mass as high as $10^{23}$ Kg. Therefore, no PBH smaller than that mass is likely to be detected.
    
A further  peculiar property of PBH decay is a lowering of the DM energy-density.  In fact, the decay effectively converts  DM into radiation, changing the total balance in the equation of state of the Universe through ages. This has a repercussion in the count of galaxies in large-scale galaxy surveys, affecting galaxy clustering, galaxy lensing and Redshift-Space Distortions (RSD)  \cite{Raccanelli:2017one}. More data concerning this will became available for instance with the LSS surveys by the SKA telescope, with the detection of individual galaxies in the radio continuum \cite{Jarvis:2015asa}: new techniques  \cite{Menard:2013aaa,Kovetz:2016hgp} allow to extract from these data the information about the evolution with respect to redshift.
If these measurements can be associated to other on PBHs, they would allow to constrain the quantum-gravity lifetime-mass scaling. The innovative aspect of such a strategy resides in the possibility to  measure a quantum-gravity phenomenon from late-universe data, as opposed to standard strategies involving data from the early universe.

Finally, the effects of the quantum-gravitational BH decay should leave an imprint in the CMB, because of the energy release by the explosion of the smallest PBHs. An ongoing effort is now addressing these effects and the analysis of the new constrains, combining different observables and considering extended PBH mass functions \cite{Bellomo:prep}.

\begin{acknowledgements} FV thanks Aur\'elien Barrau, Michael Kavic, Alvise Raccanelli, Carlo Rovelli for their collaboration on the topics exposed here. ~ The work of FV at UPV/EHU is supported by the the grant IT956-16 of the Basque Government.
\end{acknowledgements}



\end{document}